\documentclass[usenatbib]{mn2e}
\usepackage{graphicx, amssymb, aas_macros, natbib,array}
\usepackage{mathrsfs}
\usepackage{amsmath}
\usepackage{gensymb}
\usepackage{mathrsfs}
\usepackage{txfonts}
\usepackage{verbatim}

\usepackage{xcolor}

\newcommand{\mvir}{M_{\rm{vir}}}

\newcommand{\rvir}{R_{\rm{vir}}}

\newcommand{\mstar}{M_{\rm \star}}
\newcommand{\msun}{\rm \, M_{\odot}}

\newcommand{\kpc}{\rm \, kpc}

\newcommand{\mss}{{\rm m \, s}^{-2}}

\newcommand{\lstar}{L_{\star}}

\newcommand{\gobs}{g_{\rm obs}}
\newcommand{\gbar}{g_{\rm bar}}
\newcommand{\rgal}{R_{\rm gal}}
\newcommand{\mbar}{M_{\rm bar}}
\newcommand{\gdag}{g_{\rm \dagger}}

\begin{document}

\voffset=-0.6in

\vspace{-0.7cm}
\title[RAR=BTFR]
{The radial acceleration relation is a natural consequence of the baryonic Tully-Fisher relation\vspace{-0.7cm}}\author[C. Wheeler et al.]{Coral Wheeler$^1$\thanks{\vspace{-0.5cm}$\!$coral@caltech.edu}, Philip F.~Hopkins$^1$, Olivier Dor\'{e}$^{1,2}$\\
\noindent$\!\!$ $^{1}$TAPIR, Mailcode 350-17, California Institute of Technology, Pasadena, CA 91125, USA,\\
\noindent$\!\!$ $^{2}$Jet Propulsion Laboratory, 4800 Oak Grove Drive, Pasadena, CA 91109, USA
\vspace{-0.7cm}}

 \pagerange{\pageref{firstpage}--\pageref{lastpage}} 
 \pubyear{2015}
\maketitle
\vspace{-2.0cm}
\label{firstpage} 
%

\vspace{-1.0cm}
\begin{abstract} 
Galaxies covering several orders of magnitude in stellar mass and a variety of Hubble types have been shown to follow the ``Radial Acceleration Relation" (RAR), a relationship between $\gobs$, the observed circular acceleration of the galaxy, and $\gbar$, the acceleration due to the total baryonic mass of the galaxy. For accelerations above $10^{10}~\mss$, $\gobs$ traces $\gbar$, asymptoting to the 1:1 line. Below this scale, there is a break in the relation such that $\rm \gobs \sim \gbar^{1/2}$. We show that the RAR slope, scatter and the acceleration scale are all natural consequences of the well-known baryonic Tully-Fisher relation (BTFR). We further demonstrate that galaxies with a variety of baryonic and dark matter (DM) profiles and a wide range of dark halo and galaxy properties (well beyond those expected in CDM) lie on the RAR if we simply require that their rotation curves satisfy the BTFR. We explore conditions needed to break this degeneracy: sub-kpc resolved rotation curves inside of ``cored'' DM-dominated profiles and/or outside $\gg 100\,$kpc could lie on BTFR but deviate in the RAR, providing new constraints on DM.
\end{abstract}

 \begin{keywords}
 	galaxies: fundamental parameters -- galaxies: kinematics and dynamics
\vspace{-0.7cm}
 \end{keywords}

\section{Introduction}
\label{sec:intro} 

The Tully-Fisher relation \citep{Tully-Fisher1977} is a well-studied empirical relationship between the observed luminosities and rotational velocities of galaxies. While this particular form of the relation appears to break down at low masses \citep{Persic1991}, \citet{McGaugh2000} and others have shown that the ``baryonic Tully-Fisher relation'' (BTFR) between total baryonic mass ($M_{\rm bar} \equiv M_{\rm gas} + M_{\ast}$, where $M_{\ast}$ is the stellar mass and $M_{\rm gas}$ the cool/warm ISM mass of the galaxy) and circular velocity ($V_{\rm f}$ being the true, model-inferred circular velocity $V_{\rm f}^{2} \equiv R\,\partial\Phi/\partial R $ measured within the ``flat'' portion of the rotation curve)  exhibits a tighter correlation extending to dwarf masses. Using 21cm rotation curves from a large sample of galaxies with $30 \lesssim V_{\rm f} \lesssim 300\,{\rm km\,s^{-1}}$, \citet{McGaugh2005} fits $M_{\rm bar} \approx \mathscr{A}\,V_{\rm f}^{b}$ with $\mathscr{A}\approx 50 \msun km^{-4}s^4$, $b = 4$, and $\sim 0.1\,$dex lognormal scatter.  

The BTFR was historically challenging to explain within the framework of $\Lambda$CDM galaxy-formation simulations, as models which simply convert most of the baryons into stars predict a different slope ($b=3$) and normalization from that observed. However, more recent simulations accounting for stellar feedback processes have shown this removes low-angular momentum gas and suppresses star formation in low-mass halos, in a manner which dramatically improves the agreement with the BTFR \citep{Brook2012, Hopkins2014a, Crain2015, Sales2017}.

Recently, \citet{McGaugh2016} and \citet[][hereafter L17]{Lelli2017b} have described a generalization of the BTFR, the ``radial acceleration relation" (RAR). The RAR is a relationship between the observed ``gravitational acceleration'' $\gobs \equiv V_{\rm c}^2(r) / r$ (with $V_{c}$ observed at radius $r$), and the acceleration (assuming a spherical potential) due to the enclosed baryonic mass in the same radius, $\gbar \equiv G M_{\rm bar}(<r) / r^{2}$. Essentially, the RAR plots the BTFR at every point along each galaxy's rotation curve. At high acceleration ($\gbar \gtrsim \gdag \sim 10^{-10}~\mss$), $\gobs\approx\gbar$, as expected if the highest-density regions are baryon-dominated. At low acceleration, $\gobs > \gbar$ with a slope of $\gobs \propto \gbar^{1/2}$ in the outskirts of massive galaxies and dwarfs (possibly flattening further still for lower-mass dSphs; see Figs.~8-10 in L17). A fit which interpolates between both regimes is given by L17 as $\gbar/\gobs = 1 - \exp{(\sqrt{\gbar/\gdag})}$ (with a $\sim 0.1-0.15\,$dex scatter, rising towards lower masses).  

The functional form of the RAR and value of the scale $\gdag$ has led some authors \citep[][L17]{McGaugh2016, Milgrom2016} to suggest the acceleration discrepancy is due to an alternative law of gravity, specifically the \citet{Milgrom1983} form of Modified Newtonian Dynamics (MOND) which fits such a functional form and value of $\gdag$ to individual rotation-curve measurements. Others have argued the relation is a natural consequence of $\Lambda$CDM and that simulated galaxies with standard gravity+DM follow the relation \citep{Keller2007, Ludlow2017}. 

However, we show in this Letter that the RAR is not an entirely new property of galaxies, but its slope, scatter, and acceleration scale follow directly from the BTFR. We demonstrate that for a wide variety of galaxy or DM halo properties, requiring the galaxy lie on BTFR forces it to obey the RAR over its entire (measured) radial extent. We then identify the extreme systems/radii which must be measured to break this degeneracy.

\begin{table*}
	\centering
	\begin{tabular}{| c | c  c  c c c c c c| }
		\hline
		& DM Profile & Baryonic Profile & $\mvir (10^{12}\msun)$ & $\rvir (\kpc)$ &  $\rm c$ & $\mbar (10^{10}\msun)$ & $\rgal(\kpc)$ & $\rm BTF \, norm \,  coeff$ \\ 
		\hline
		FID & NFW & EXP & $1.0$ & $260.1$  & $8.3$ & $3.6$ & $5.7$ & $1$ \\ 
        \hline
		Rgal--X & NFW & EXP   & $1.0$ & $260.1$ & $8.3$ & $3.6$  & $X$  & $1$\\
		Mbar--X & NFW & EXP   & $1.0$ & $260.1$ & $8.3$ & $X/1e10$ & $5.7$ & $1$  \\
		MbRg--X--Y & NFW & EXP   & $1.0$ & $260.1$ & $8.3$ & $X/1e10$ & $Y$ & $1$  \\
        \hline
		C--X & NFW & EXP & $1.0$ & $260.1$  & $X$ & $3.6$ & $5.7$ & $1$\\  
		Rvir--X & NFW & EXP & $1.0$ & $X$  & $8.3$ & $3.6$ & $5.7$ & $1$\\  
		Mvir--X & NFW & EXP & $X/1e12$ & $260.1$  & $8.3$ & $3.6$ & $5.7$ & $1$\\  
		MvRv--X--Y & NFW & EXP & $X/1e12$ & $Y$  & $8.3$ & $3.6$ & $5.7$ & $1$\\  
        \hline
		MmRrC--2e9 & NFW & EXP & $0.002$ & $32.8$  & $15.6$ & $0.0007$ & $0.1$ & $1$\\  
		MmRrC--5e10 & NFW & EXP & $0.05$ & $95.9$  & $11.3$ & $0.03$ & $1.3$ & $1$\\  
		MmRrC--2.5e13 & NFW & EXP & $25.0$ & $760.5$  & $6.0$ & $13.1$ & $9.5$ & $1$\\  
		MmRrC--2.5e14 & NFW & EXP & $250.0$ & $1533.6$  & $4.8$ & $21.3$ & $12.5$ & $1$\\  
		\hline
		dm\_CIS & CIS & EXP & $1.0$ & $260.1$  & $16.6$ & $3.6$ & $5.7$ & $1$\\
		dm\_CNFW & CNFW & EXP & $1.0$ & $260.1$  & $8.3$ & $3.6$ & $5.7$ & $1$\\
		dm\_HERN & HERN & EXP & $1.0$ & $260.1$  & $8.3$ & $3.6$ & $5.7$ & $1$\\
		dm\_SIN & SIN& EXP & $1.0$ & $260.1$  & $4.2$ & $3.6$ & $5.7$ & $1$\\		
		\hline
		bar--HERN & NFW & HERN & $1.0$ & $260.1$  & $16.6$ & $3.6$ & $4.0$ & $1$\\  
		bar--SIS & NFW & SIS & $1.0$ & $260.1$  & $8.3$ & $3.6$ & $22.8$ & $1$\\ 
		bar--CNFW & NFW & CNFW & $1.0$ & $260.1$  & $8.3$ & $3.6$ & $1.7$ & $1$\\  
		bar--SIN & NFW & SIN & $1.0$ & $260.1$  & $8.3$ & $216.0$ & $40.0$ & $1$\\  
		\hline
		BTF--X & NFW & EXP & $1.0$ & $260.1$  & $8.3$ & $3.6$ & $5.7$  & $X$\\  
		\hline
	\end{tabular}
	\label{tab:table} 
	\caption{Summary of models used in Fig.~\ref{fig:f1}. Fiducial (FID) is an $L_{\star}$ galaxy in a DM halo with an NFW DM and exponential (EXP) baryonic profile, with virial mass $M_{\rm vir}$ and radius $R_{\rm vir}$, concentration $c$, baryonic mass $M_{\rm bar}$ and half-mass radius $R_{\rm gal}$, as labeled. The profile is normalized to have $V_{\rm f}$ a multiple (``BTF norm coeff'') of the observed BTFR. Where one or two parameters are varied, we label ``Name--$X$--$Y$'' (e.g.\ $R_{\rm gal}--X$ keeps all values fixed as shown, but varies $R_{\rm gal}=X\,{\rm kpc}$ according to the label in Fig.~\ref{fig:f1}).\vspace{-0.4cm}}
\end{table*}

\vspace{-0.7cm}
\section{Basic Scalings}
\label{sec:analytic} 

First, we consider simple scalings which demonstrate that the RAR scalings follow from the BTFR, independent of the physical origins of the ``anomalous acceleration.'' At high acceleration (exclusively the inner regions of more massive galaxies) the RAR has $\gobs\approx\gbar$, consistent with gravity being Newtonian and from baryons alone. This follows trivially if high-acceleration (high-density) regions are baryon-dominated, i.e., DM cannot become arbitrarily dense. 

At low acceleration (large radii) the RAR asymptotically approaches $\gobs \approx \sqrt{\gbar\,\gdag}$. But consider: the observed BTFR is $M_{\rm bar} = \mathscr{A}\,V_{\rm f}^{4}$, where $V_{\rm f}$ is (by definition) measured on the {\em flat} part of the rotation curve so $V_{c}(r)\approx V_{\rm f}$ over a large range of $r$. If we are outside the range where $M_{\rm bar}(<r) \sim M_{\rm bar}$ (i.e., the baryonic mass is starting to converge), then $\gbar \equiv G M_{\rm bar}(<r)/r^{2} \approx G M_{\rm bar}/r^{2} = G (\mathscr{A}\,V_{\rm f}^{4}) / r^{2} = G \mathscr{A}\,V_{\rm c}^{4}(r) / r^{2} = G \mathscr{A}\,(V_{c}^{2}[r]/r)^{2} = (G \mathscr{A})\,\gobs^{2}$. So we obtain exactly $\gobs = \sqrt{\gbar\,\gdag}$ where $\gdag = (G \mathscr{A})^{-1} \approx 1.5\times10^{-10}~\mss$ (using the measured $\mathscr{A}$). 

This also explains the scatter of the RAR. In the baryon-dominated region all scatter should be measurement error or physical noise (e.g.\ non-spherical corrections). At low acceleration if we assume a galaxy deviates from the BTFR by a factor $\delta$ (mass $\mbar = \mathscr{A} V_c^4 (1+\delta)$), then repeating the above gives $\gobs=(1+\delta)^{-1/2}\,\sqrt{\gbar\,\gdag}$, so the logarithmic scatter in the RAR is that in the BTFR, reduced by a factor $1/2$ (from the square root).

\vspace{-0.5cm}
\section{Example \&\ Constraints on Mass Profiles}
\label{sec:results} 

\begin{figure*}
\begin{tabular}{cc}
\vspace{-0.25cm}
\hspace{-0.15cm}\includegraphics[width=0.49\textwidth]{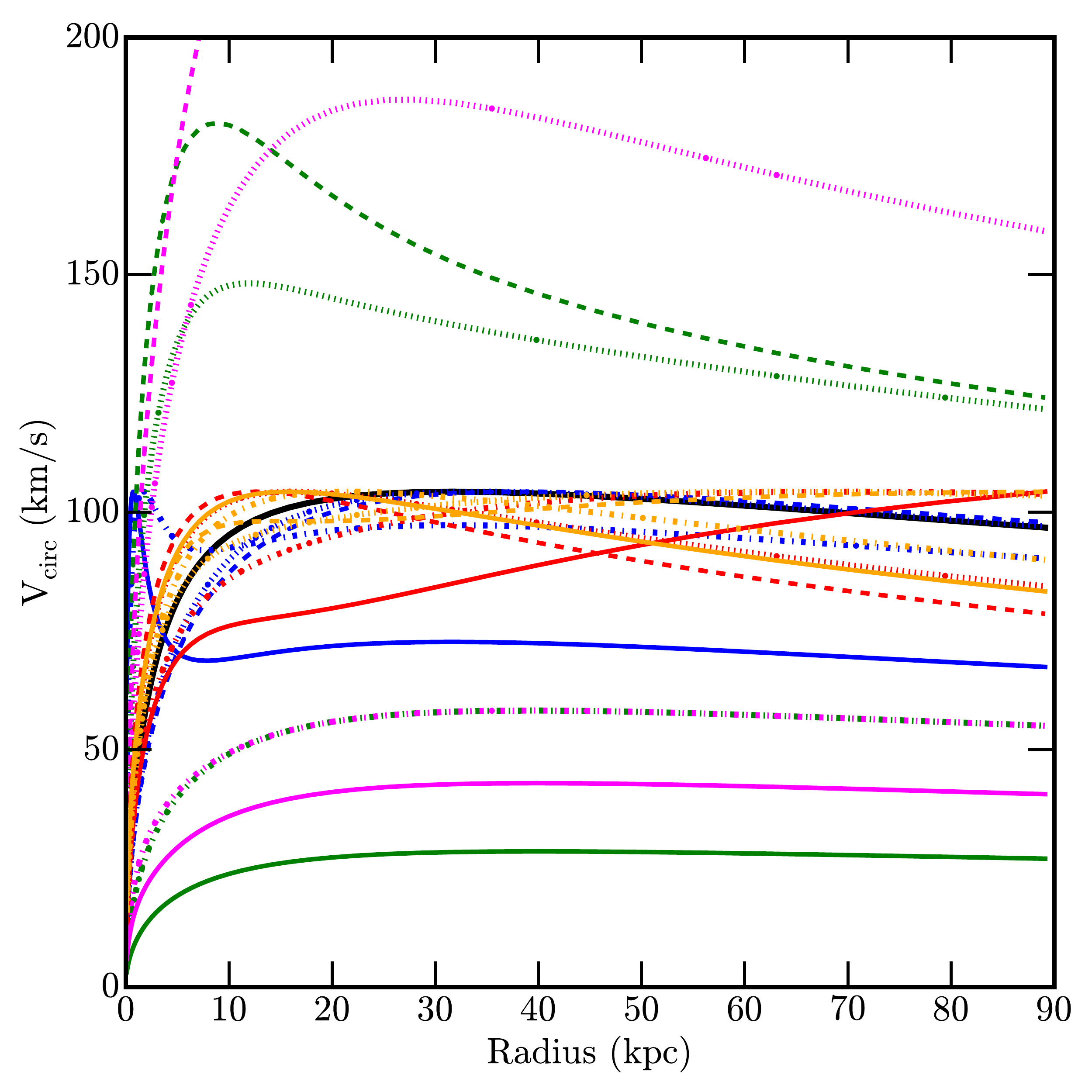} & 
\hspace{-0.45cm}\includegraphics[width=0.49\textwidth]{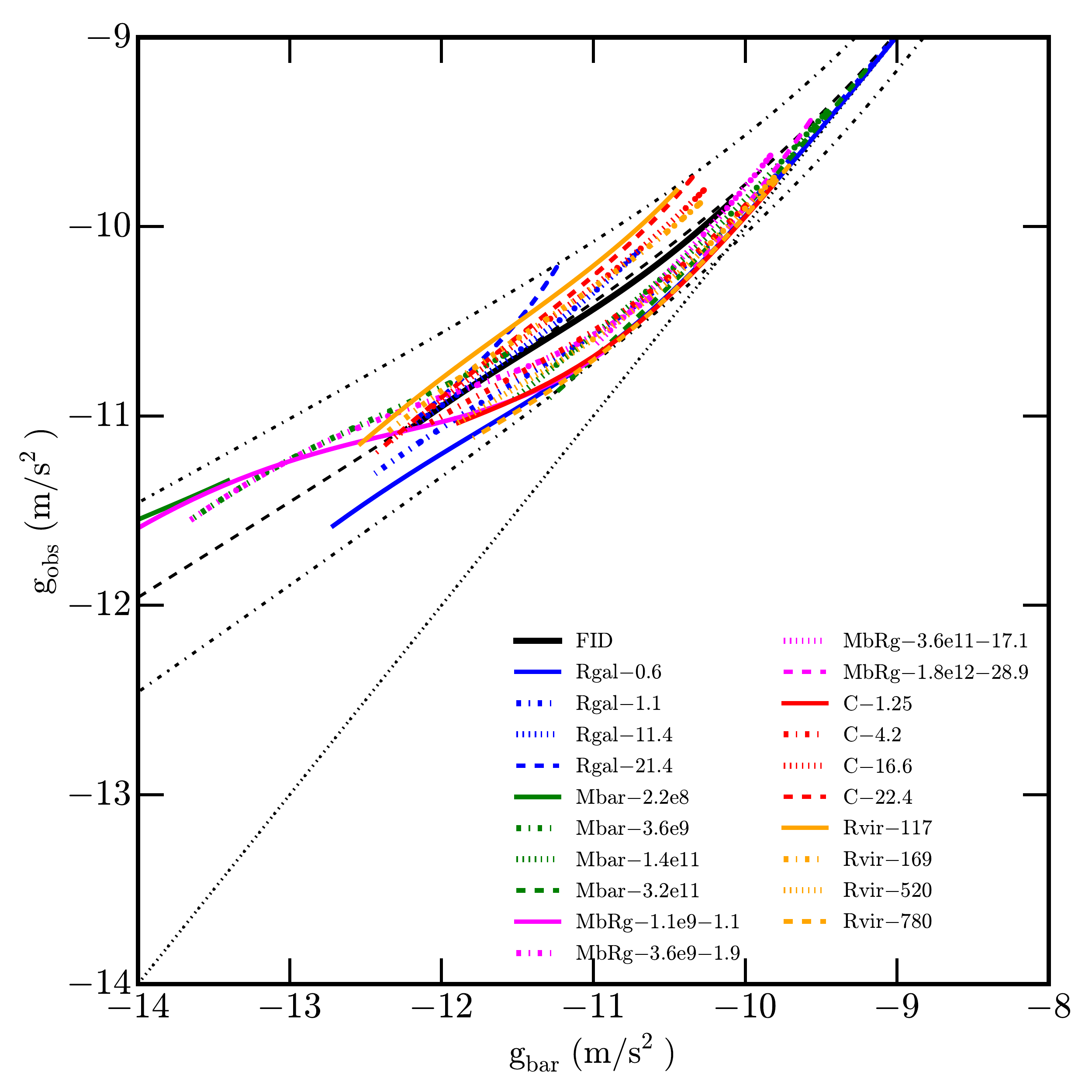} \\
\vspace{-0.25cm}
\hspace{-0.15cm}\includegraphics[width=0.49\textwidth]{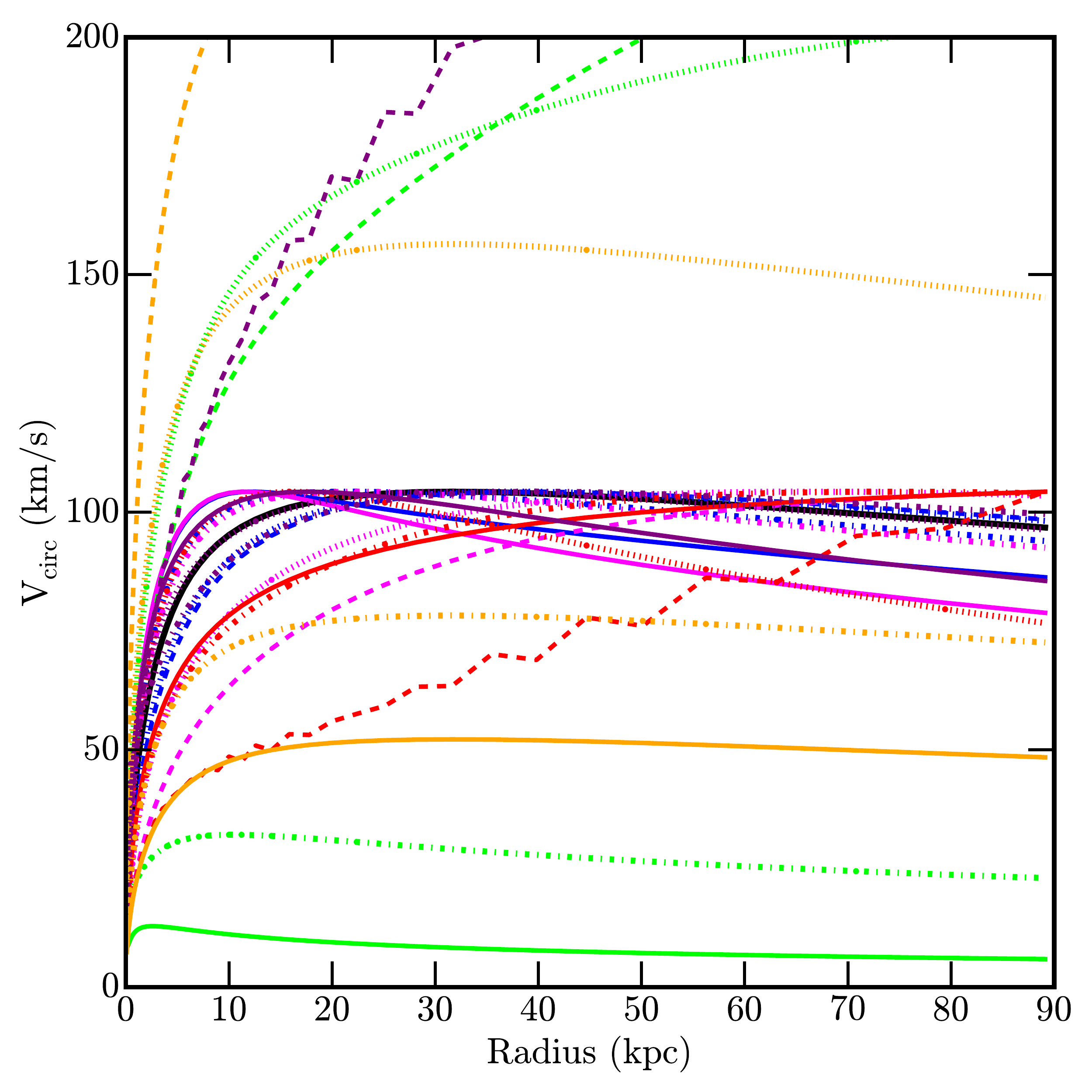} & 
\hspace{-0.45cm}\includegraphics[width=0.49\textwidth]{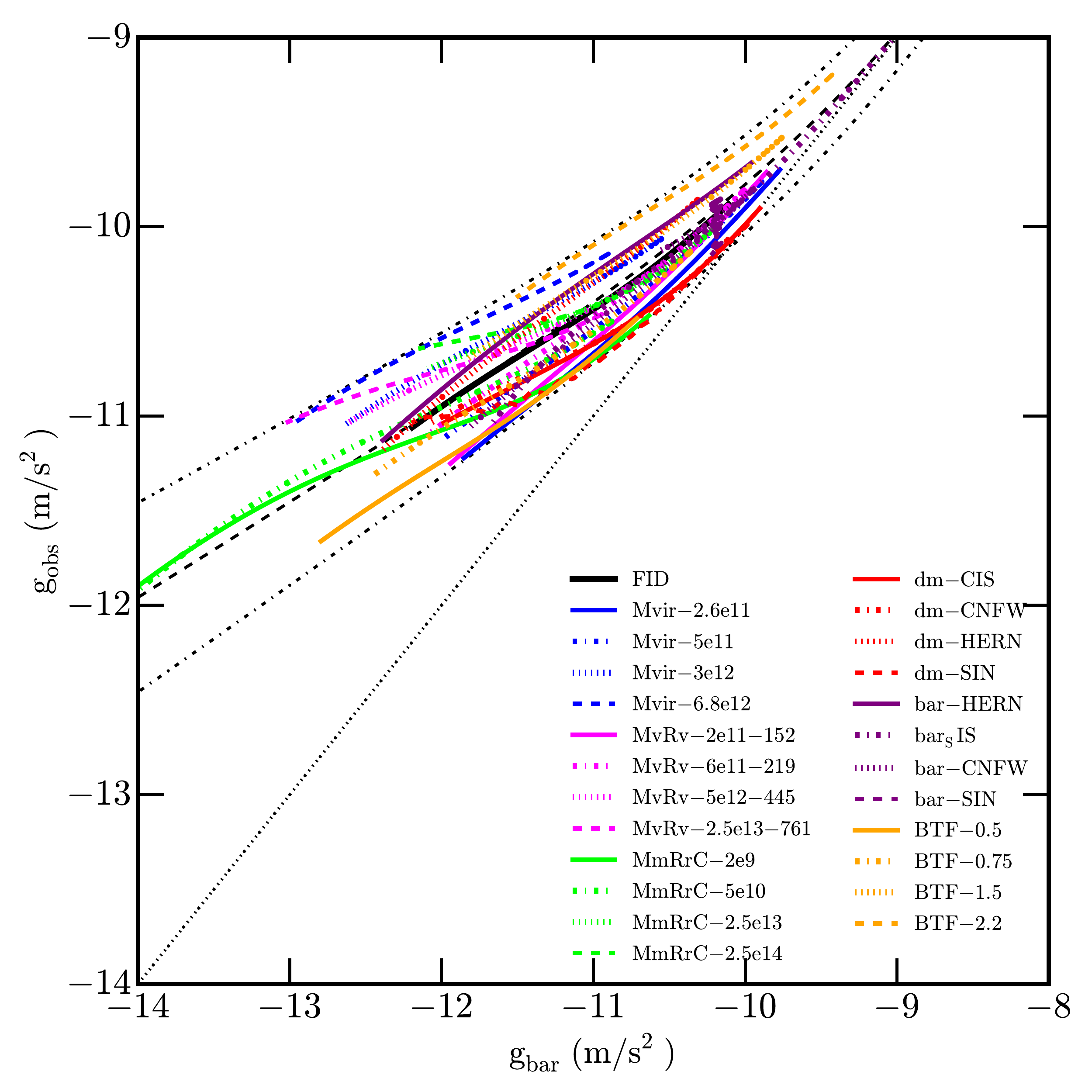} 
\end{tabular}
\vspace{-0.25cm}
\caption{{\em Left}: Rotation curves for the different baryon+DM models listed in Table~\ref{tab:table}. Upper and lower panels plot different subsets varying different parameters, as labeled.
{\em Right}: Location of each point in the rotation curve at {\em left}, plotted on the RAR throughout the radial range $0.1 \kpc< \it r < \rm 89 \kpc$. We compare the observed best fit (dashed black lines) and $95\%$ ($\pm2\,\sigma$) inclusion interval (dash-dotted black) and $\gobs=\gbar$ (dotted black). Despite enormous (often unphysical) variations in profile shapes, sizes, baryon-to-DM masses, and more, all lie within the observed range. The only condition we impose is that the model profile lie on the observed BTFR.
\label{fig:f1}
} 
\end{figure*}

Fig.~\ref{fig:f1} demonstrates more rigorously that galaxies with a variety of properties will fall within observed scatter on the RAR if they also satisfy the BTFR. Consider simple two-component (``baryon''+``dark'') models: for example take a (spherical) DM halo with an \citep[][NFW]{NFW1997} profile $\rho_{\rm DM}\propto r^{-1}\,(1 + r/r_{s})^{-2}$, with scale radius $r_{s} = r_{\rm vir}/c$ ($r_{\rm vir}$ the virial radius and $c$ the concentration), normalized by the total mass $M_{\rm vir}$ inside $r_{\rm vir}$. Add a baryonic component with an exponential surface density profile ($\Sigma(R) \propto \exp{(-R/\rgal)}$), with total mass $M_{\rm bar}$. 

If desired, ``typical'' galaxy properties can be estimated from observed scalings.\footnote{Choosing $\mvir$, we can use the abundance-matching relation from \citet{Garrison-Kimmel2014a} (identical to \citealt{Behroozi2013a} at high mass) to determine $\mstar$. Then $\log_{10}(M_{\rm HI}/\mstar) = -0.43\log_{10}(\mstar/\msun) + 3.75$ \citep{Papastergis2012} gives $\mbar$ (using the same He-correction $\mbar = \mstar + 1.33\,M_{\rm HI}$ as used in the RAR observations themselves). For massive galaxies ($\mstar>10^{9}\,\msun$) $\rgal = 0.13\,{\rm kpc}\,\alpha\,(\mstar/\msun)^{0.14}\,(1 + \mstar/1.4e11\,\msun)$ \citep{Lange2015}, and for dwarfs ($\mstar<10^{9}\,\msun$) $\rgal = 3.1\,{\rm kpc}\,\alpha\,(\mstar/10^{9}\,\msun)^{0.405}$ \citep{DiCintio2016}, where $\alpha=0.6,\,1.2$ for stars/gas, and we take the mass-average for $\mbar$. At fixed $\mvir$, $\rvir$ follows immediately from cosmology (here ``concordance'' $\Lambda$CDM) and $c = 8.3\,(\mvir/10^{12}\,\msun)^{-0.1}$ \citep{Dutton2014}. For an $\sim L_{\ast}$ ($10^{12}\,\msun$) halo this gives the ``fiducial'' parameters in Table~\ref{tab:table}.} We will use these as a starting point but in fact treat these parameters as essentially free and vary them in turn as shown in Table~\ref{tab:table} \&\ Fig.~\ref{fig:f1}. Starting from a fiducial case, we vary $\rgal$, $\mbar$, $\mvir$, $\rvir$, and $c$ each independently keeping all other properties fixed, so e.g.\ our models do {\em not} lie on any observed relations; we then co-vary $\mbar$ and $\rgal$ along the observed relation, then $\mvir$ and $\rvir$ along the standard cosmological relation; then co-vary $\mvir$, $\rvir$, $c$, $\mbar$, $\rgal$ all along the observed scaling laws together. Next we consider varying the mass profile of DM or baryons; e.g.\ for DM modifying NFW to a cored-pseudo isothermal (CIS; $\rho \propto (1 + r^{2}/r_{s}^{2})^{-1}$), isothermal (SIS; $\rho \propto r^{-2}$), cored-NFW (CNFW; $\rho \propto (1+r/r_{s})^{-3}$), or \citet{Hernquist1990} (HERN; $\rho \propto r^{-1}\,(1+r/r_{s})^{-3}$) profile (for baryons we consider the same set with $r/r_{s} \rightarrow r/\rgal$). For illustrative purposes we also consider an intentionally un-physical model (SIN) where the density varies wildly and repeatedly with radius between zero and some maximum value: $\rho \propto r^{-1}\,\sin^{2}(\ln{4\pi(r/r_{s})})$.

For each model, we then calculate $V_{c}^{2}(r) \approx G M_{\rm enc}(<r)/r$ from the enclosed mass ($M_{\rm enc}$) profile, where the full rotation curve (calculated for $0.1 \kpc < \it r < \rm 89 \kpc$) is a combination of DM+baryonic components: $V_{c}^{2}(r) = V_{c,\,{\rm DM}}^{2}(r) + V_{c\,{\rm bar}}^{2}(r)$. We then apply the crucial step: we {\em require} that the rotation curve obey the BTFR. Specifically, we require that $V_{\rm f}$ (which we approximate as the maximum of the total $V_{c}$, or value at $90\,$kpc if it continues rising at larger radii) is equal to $(M_{\rm bar}/\mathscr{A})^{1/4}$ (with the measured $\mathscr{A}$ from L17).\footnote{Numerically, we calculate $V_{\rm f}^{0}$ from the initial profile ``guess,'' noting that if we re-normalize the {\em total} DM+baryonic masses by a uniform factor $f^{2}$, we re-normalize $V_{\rm f}^{0}$ by $f$, so setting $V_{\rm f} = f\,V_{\rm f}^{0} = V_{\rm obs} = (\mbar/\mathscr{A})^{1/4}$ we obtain $f^{2} = \mbar^{0}/\mathcal{A}\,(V_{\rm f}^{0})^{4}$.} This essentially removes one degree of freedom from the models (e.g.\ the absolute mass scale at a given $V_{\rm f}$ is ``fixed'' by this assumption), but the sizes of baryons and DM, their mass ratios, and their profile shapes (essentially all parameters that determine the {\em shape} of the rotation curve) remain freely-varied. Fig.~\ref{fig:f1} shows these models all remain inside the observed range.

For the models with varied parameters, we intentionally vary them until we find extremal values where the predicted RAR lies at the ``edge'' of the observed $95\%$ inclusion contour. For example, our ``Mbar\_'' (``Rgal\_'') series tells us that $\sim 95\%$ of galaxies with $\mvir=10^{12}~\msun$ must have $2.2\times10^{8} < \mbar/\msun < 3.2\times10^{11}$ ($0.6<\rgal/{\rm kpc}<21.4$). The actual observed scatter in the $M_{\rm gal}-M_{\rm halo}$ relation at this mass is just $\sim 0.1\,$dex, much smaller than this range (likewise for sizes). Similarly the allowed $\mvir$, $\rvir$, and $c$ range at a given $\mbar$ range is much larger than obtained in cosmological simulations (at $\sim L_{\ast}$, $2\times10^{11}< \mvir/\msun < 2.5\times10^{13}$, $117 < \rvir/{\rm kpc} < 780$, $1.2 < c < 22.4$). We can vary the mass profiles (of DM and/or baryons) well beyond those observed in galaxies or cosmological simulations (with inner slopes from $\rho \propto 1/r^{0-2}$ and outer from $\rho \propto 1/r^{1-4}$). Of course, these ``allowed'' parameter ranges should be taken as maximal limits, because we have {\em imposed} that the systems lie on the BTFR, i.e. by re-normalizing the curves, we de-facto adjust the parameters themselves and so the resulting range of values listed above will be much larger than the true physical limits. We have plotted the acceleration over quite a large radial extent ($0.1 \kpc < \it r < \rm 89 \kpc$). If we were to increase the radial range over which we measure the acceleration, the ``allowed" ranges would decrease. What we have shown is that the BTFR is much more important for the RAR than any other varied galaxy property here. In other words, other correlations like size-mass, concentration-mass, or $M_{\rm gal}-M_{\rm halo}$, and mass profile shapes, do not (within reasonable physical values) play a significant role in the RAR, {\rm provided} BTFR.

We see this directly by simply varying the location of models with respect to the BTFR: our ``BTF\_'' systematically changes the normalization $\mathscr{A}\rightarrow f\,\mathscr{A}_{\rm obs}$ of the BTFR used to normalize the rotation curves. As expected this directly shifts the position in the RAR; moreover the $95\%$ range observed in the RAR corresponds almost exactly to the $95\%$ range in the observed BTFR. 

\begin{figure}
	\includegraphics[width=0.49\textwidth]{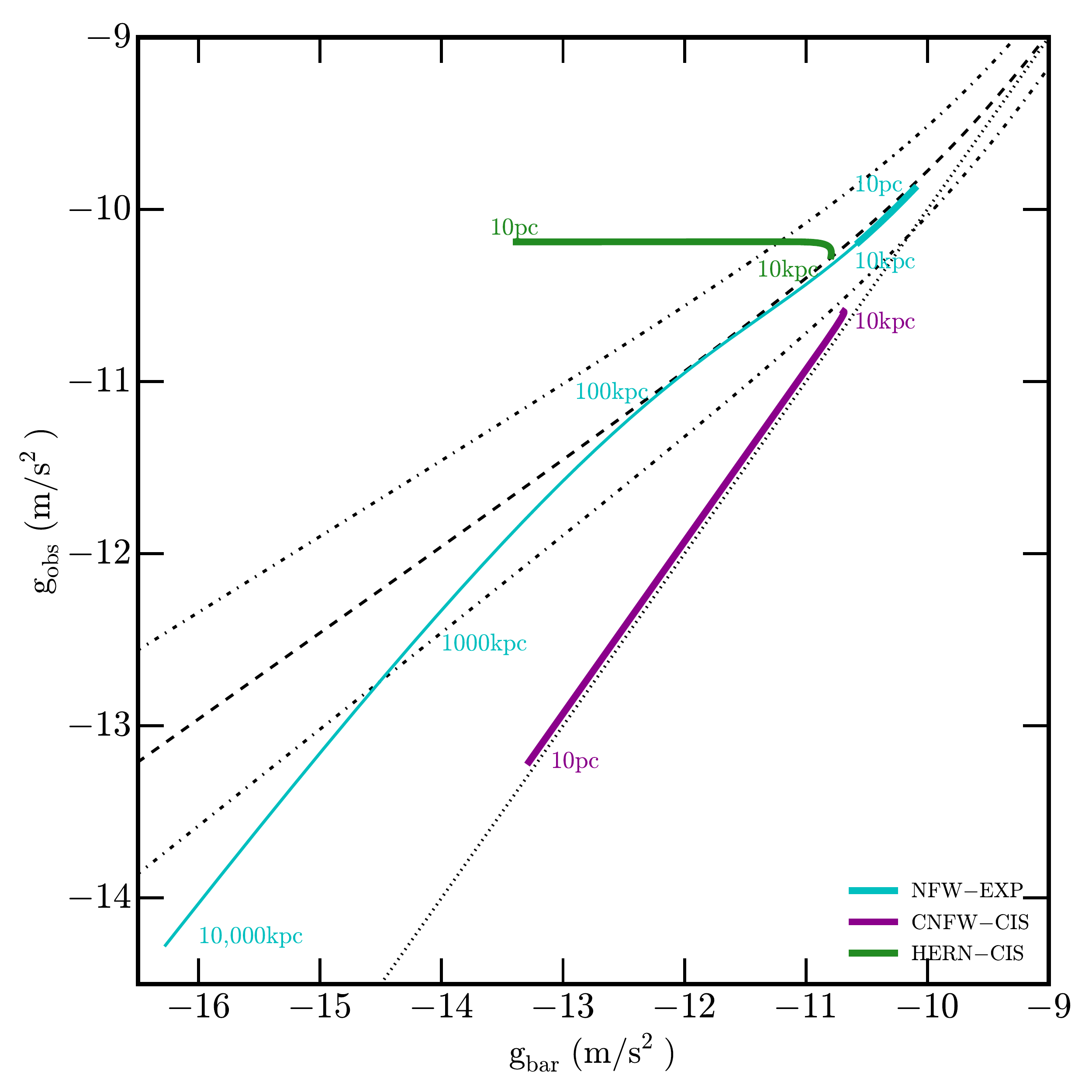} 
	\caption{Radial acceleration relation for our fiducial model with different baryonic and non-baryonic density profiles chosen to highlight the behavior at small and large radii. The observed best fit (dashed black) and $95\%$ ($\pm2\,\sigma$) inclusion interval (dash-dotted black), and $\gobs=\gbar$ (black dotted) are shown for comparison. The thick lines show the RAR for $0.01 \kpc< \it r < \rm 10 \kpc$ for the NFW (EXP), CNFW (CIS) and HERN (CIS) dark matter (baryonic) profiles (cyan, dark magenta and green lines respectively), while the thin line shows the NFW (EXP) profile for $100 \kpc< \it r < \rm 10^4 \kpc$. The numbers highlight various radii at which the points on the $\gobs-\gbar$ relation are calculated. At large $r$, the NFW (EXP) model (cyan line), and in fact all models with exponential baryonic profiles and dark profiles that fall off as $r\gtrsim 3$ eventually fall off the RAR at large radius. At small $r$, those same models do not deviate from the RAR because both profiles approach constant values as $r\rightarrow 0$. Profiles that have cores in both the dark and baryonic component (similar to the CNFW-CIS, thick dark magenta line) deviate from the RAR by tracing the 1:1 line because both observed and baryonic acceleration profiles go $\sim r$ at small radii. Models with cored baryonic profiles and cuspy dark profiles (e.g. HERN-CIS, thick green line) have $\gobs \rightarrow const$ at the center while $\gbar \sim r$.
		\label{fig:f2}
 	} 
\end{figure}

\vspace{-0.5cm}
\section{Breaking the RAR}
\label{sec:breaking}

{Can} galaxies deviate from the RAR, in a manner which provides new information beyond their location on the BTFR? At high accelerations $\gg \gdag$ (where $\gobs\approx\gbar$), deviations from the RAR would require a ``dark'' component dominate the acceleration over baryons; this in turn requires a minimum DM acceleration $G M_{\rm enc}^{\rm DM}(<r)/r^{2} \approx \pi\,G \langle\Sigma_{\rm DM}\rangle \gg \gdag$, which would require DM collapse to surface densities $M_{\rm enc}/(\pi\,R^{2}) \gg 500\,\msun\,{\rm pc^{-2}}$ (denser than typical GMCs) on $\sim$kpc scales. This is ruled out by lensing and other constraints (and is not expected for standard collisionless particle-CDM candidates). 

At intermediate accelerations, $\sim \gdag$, we are very close to the radii where the BTFR relation is measured (each point on BTFR is a point on RAR at about this $\gobs$) -- so any deviation in the RAR is equivalent to a deviation from the BTFR. 

At weak accelerations $\ll \gdag$ (where $\gobs \approx \sqrt{\gbar\,\gdag}$), the fact that the RAR plots all radial points makes it possible that a galaxy could lie on BTFR and still deviate in the RAR. Recall, {\em at the radius $r_{0}$ where the BTFR is measured, the galaxy must also lie on the RAR} (i.e.\ $\gobs(r_{0}) = \langle \gobs(\gbar[r_{0}]) \rangle \approx \sqrt{\gbar(r_{0})\,\gdag}$). However as we extrapolate to $r \ll r_{0}$ or $r \gg r_{0}$, a deviation can occur. Noting $\gobs \approx G M_{\rm enc}(<r)/r^{2} = (M_{\rm enc}[<r]/M_{\rm enc}[<r_{0}])\,(r_{0}/r)^{2}\,\gobs(r_{0})$ (similar for $\gbar$) we can write the ratio of $\gobs$ at some radius $r$ to that expected if the galaxy lay exactly on the RAR ($\langle \gobs(r) \rangle = \sqrt{\gbar(r)\,\gdag}$) as $\gobs/\langle \gobs\rangle = [M_{\rm enc}(r)/M_{\rm enc}(r_{0})]\,[M_{\rm bar}(r_{0})/M_{\rm bar}(r)]^{1/2}\,(r_{0}/r)$. 

Now consider $r\ll r_{0}$: this can be seen in the thick lines of Figure \ref{fig:f2}, where the RAR is plotted for a model galaxy with the fiducial parameters except for varying density profiles over the radial range $0.01 \kpc< \it r < \rm 10 \kpc$. If the system becomes very dense at small radii (i.e.\ has a steep cusp), $\gobs$ will increase, giving the high-acceleration regime above. NFW-type DM profiles ($\rho\propto r^{-1}$) and exponential-type (constant inner-surface-brightness) baryonic profiles (thick cyan line) produce zero deviation ($\gobs\approx\langle \gobs\rangle$) as $r\rightarrow 0$. Deviations are maximized for flatter $\rho$ profiles (where rotation curves rise steeply as possible), so consider the case where both DM+baryons have a core ($\rho\propto r^{0}$) out to some radius $r_{\rm core}$ (where the system is close to BTFR);\footnote{Note the scaling above within the core} then $\gobs/\langle \gobs \rangle \approx (r/r_{\rm core})^{1/2}$. So to deviate by more than a factor of $2 (\approx \pm 2\,\sigma$, at weak accelerations) from the median observed RAR, one must measure the RAR at $r \lesssim r_{\rm core}/4$. This can be seen in the thick dark magenta line in Figure \ref{fig:f2}, where both the observed and baryonic accelerations $\sim r$, at low radii, and thus trace the 1:1 line. Models with cored baryonic profiles and cuspy dark profiles (e.g. HERN-CIS, thick green line) have $\gobs \rightarrow const$ at the center while $\gbar \sim r$. Although Figure \ref{fig:f2} shows the profiles for an $\lstar$ galaxy, because dark matter cores are observed only in low-mass galaxies, observationally detecting this deviation likely requires sub-kpc resolution of the rotation curve. The existence of cores may explain why there appears to be a flattening of the relation in observed dwarfs (L17, their Figure 12). It is interesting to note that the largest deviations occur for $\rm 2 \lesssim log(M_{tot}/M_{bar}) \lesssim 3$ -- approximately the ratio of halo to baryonic (stellar) mass for which core formation is likely to be most efficient \citep{DiCintio2014a}.

Conversely, consider $r \gg r_{0}$: at large radii $M_{\rm bar}(<r)\approx M_{\rm bar}$ so $\gobs/\langle \gobs \rangle \approx [M_{\rm enc}(r)/M_{\rm enc}(r_{0})]\,(r_{0}/r) = V_{c}^{2}(r)/V_{c}^{2}(r_{0})$, and deviations are maximized if the total density profile (dominated by DM) is falling steeper-than-isothermal (so rotation curves decline, instead of being flat). This can be seen in the thin cyan line of Figure \ref{fig:f2}, where the RAR is plotted for a model galaxy with the fiducial parameters and the NFW (EXP) profile for dark matter (baryonic component) over the radial range $100 \kpc< \it r < \rm 10^4 \kpc$.  Galaxies with dark matter density profiles that fall off at least as steeply as $r^{-3}$ and exponential baryonic profiles have $\rho\propto r^{-3}$, so $\gobs/\langle \gobs \rangle \propto r^{-1}\,(1+\ln{(r/r_{s})})$ deviate progressively from the RAR at larger $r$. However, the NFW $V_{c}$ falls very slowly: for $r_{0}\sim 0.5-2\,r_{s}$ (where $V_{c}$ is flat), deviation from the RAR by a factor $>2$ requires going to $r \gtrsim 20\,r_{s} = 20\,\rvir/c \sim 2\,\rvir$. Given the extensive radial range ($0.1 \kpc < \it r < \rm 89 \kpc$) plotted in Figure \ref{fig:f1}, it is remarkable that the curves all fall within $\pm2\sigma$ of the observed RAR simply by requiring them to lie on BTFR at a single radius.

\vspace{-0.5cm}
\section{Discussion \&\ Conclusions}
\label{sec:conclusion}

We have shown that the observed shape/slope, normalization, and scatter of the RAR derives directly from the observed slope, normalization, and scatter of the BTFR. The high-acceleration portion of the relation, where $\gobs\approx\gbar$, merely tracks baryon-dominated regions. The low-acceleration portion follows directly from the observed relation between $V_{c}$ and $\mbar$ (the BTFR): although the RAR includes each radial point separately (i.e.\ $V_{c}(r)$, not just $V_{\rm f}$), the fact that circular velocity is (by definition) smooth (since it depends the integral of enclosed mass) means that if one requires galaxies lie on the BTFR, they automatically produce the RAR from scales $\sim 1-100\,$kpc, even if baryonic or DM mass profiles or mass ratios vary wildly. The scatter in the RAR is apparently reduced from the BTFR because the parameters plotted are the square root of the BTFR parameters. We do not speculate on the physical origin of the ``acceleration scale'' $\gdag$ that separates these, but show that it is mathematically equivalent to the BTFR normalization ($\gdag = (G \mathscr{A})^{-1}$), and its seemingly universal value is merely a reflection of the low scatter in the BTFR (which sets the scatter in the normalization of $\gobs(\gbar)$). 

Previous studies \citep{DiCintio2016,Navarro2017} argued that galaxies which obey the observed scaling relations between $\mbar$, $\rgal$, profile shape, and $\mvir$, in ``standard'' $\Lambda$CDM NFW-type DM halos, naturally reproduce the RAR. Others have shown that some cosmological $\Lambda$CDM galaxy formation simulations reproduce the RAR \citep[e.g.][]{Ludlow2017}. Our study shows that these models ``work'' because their galaxies obey BTFR; moreover we consider a much broader range of parameters and show that many ``non-standard'' galaxy or DM properties would also agree with the observed RAR, if they obeyed the same BTFR. To the extent that some models {\em fail} to reproduce the RAR \citep[e.g.][others in]{Ludlow2017}, it can be directly identified with a failure to reproduce the BTFR. 

We do show it is possible for a galaxy to lie on BTFR, but deviate from the RAR at sufficiently low accelerations (high-acceleration deviations would require non-standard DM that could reach very high densities). This can occur either (1) in the very centers of cored, DM-dominated galaxies, or (2) at very large radii in NFW halos. However, if the BTFR still applies around the effective radius, this requires measurements of the rotation curve either at very small radii (sub-kpc, $r \lesssim r_{\rm core}/4$) or {\em very} large radii ($\gtrsim 2\,\rvir$), respectively. 

Of course, if the galaxy does not lie on the extrapolated BTFR (at the measured radii), it will not lie on the same RAR. There is likely already evidence for this in the data: L17 (Figs.~10-11) show that their lowest-mass dSph sample deviate significantly (with constant $\gobs \approx 10^{-11}\,{\rm m\,s^{-2}}$ independent of $\gbar$). For most of these systems, the measured $V_{c}$ lies well within the rising portion of the rotation curve if an NFW halo is assumed; for a DM-dominated NFW halo at $r\ll r_{s}$, $\gobs = G M_{\rm enc}(<r) / r^{2} \approx G \mvir / [r_{s}^{2}\,2\,\{\ln{(1+c)} -c/(1+c) \} ] \approx 1.4\times10^{-11}\,(\mvir/10^{9}\,\msun)^{0.18}\,{\rm m\,s^{-2}}$ (where the latter uses the observed scalings from \S\ref{sec:results}; note $\mvir^{0.18}\propto M_{\ast}^{0.07}$ from the same scalings at these masses) -- this agrees surprisingly well with the observed relation.

\vspace{-0.5cm}
\section*{Acknowledgments} 
\vspace{-0.05cm}
CW is supported by the Lee A. DuBridge Postdoctoral Scholarship in Astrophysics. Support for PFH was provided by an Alfred P. Sloan Research Fellowship, NSF Collaborative Research Grant \#1715847 and CAREER grant \#1455342. Part of the research was carried out at the Jet Propulsion Laboratory, California Institute of Technology, under a contract with the National Aeronautics and Space Administration.

\label{lastpage}
\end{document}